# A deep spatio-temporal attention model of dynamic functional network connectivity shows sensitivity to Alzheimer's in asymptomatic individuals

Yuxiang Wei[1,2], Anees Abrol[1,2], James Lah[3], Deqiang Qiu[3], Vince D. Calhoun[1,2]
[1]Center for Translational Research in Neuroimaging and Data Science (TreNDS), Atlanta, USA
[2]Georgia Institute of Technology, Atlanta, USA
[3]Emory University, Atlanta, USA

*Abstract*—Alzheimer's disease (AD) progresses from asymptomatic changes to clinical symptoms, emphasizing the importance of early detection for proper treatment. Functional magnetic resonance imaging (fMRI), particularly dynamic functional network connectivity (dFNC), has emerged as an important biomarker for AD. Nevertheless, studies probing at-risk subjects in the pre-symptomatic stage using dFNC are limited. To identify at-risk subjects and understand alterations of dFNC in different stages, we leverage deep learning advancements and introduce a transformer-convolution framework for predicting at-risk subjects based on dFNC, incorporating spatial-temporal self-attention to capture brain network dependencies and temporal dynamics. Our model significantly outperforms other popular machine learning methods. By analyzing individuals with diagnosed AD and mild cognitive impairment (MCI), we studied the AD progression and observed a higher similarity between MCI and asymptomatic AD. The interpretable analysis highlights the cognitive-control network's diagnostic importance, with the model focusing on intra-visual domain dFNC when predicting asymptomatic AD subjects.

## I. INTRODUCTION

Alzheimer's disease (AD), a prevalent form of dementia, is a progressive neurological syndrome characterized by memory loss and cognitive deficit [1]. Due to its irreversible nature, the timely identification of AD holds significant clinical importance, enabling the implementation of interventions to slow its progression [2]. More recently, growing efforts have shifted towards the asymptomatic stage of AD, where individuals exhibit no sign of functional or cognitive decline but have positive cerebrospinal fluid (CSF) biomarker. Exploring AD at the early stage has become a focal point, prompting numerous efforts to understand its underlying dynamics. Several pathological biomarkers, including amyloid-β peptide ($A\beta_{42}$) and total tau protein (tTau), have been identified as pivotal prodromal risk factors for asymptomatic AD [3].

Functional magnetic resonance imaging (fMRI) has emerged as a popular imaging modality for the non-invasive examination of functional connectivity (FC) coupling among brain regions, providing insights into brain changes throughout disease progression [4]. Functional network connectivity (FNC) is an extension of this concept that focuses on estimating temporal coupling among data-driven brain networks [5]. A time-resolved extension, dynamic FNC (dFNC) captures recurring temporal patterns of FNC, enabling an understanding of temporal variability [6]. Notably, studies have demonstrated a strong correlation between dFNC and AD. For example, Sendi et al. [7] studied both static and dynamic FNC and found increasing connectivity in the middle frontal gyrus as AD progressed.

Recent literature has leveraged machine learning methods for AD diagnosis based on FNC. Millar et al. [8] applied a Gaussian regression model to select brain networks and used a linear regression model to predict at-risk subjects. Shen et al., [9] employed a decision-tree-based method called Hollow Tree Super to predict patients with subject cognitive decline. Alorf et al. [10] utilized a stacked sparse auto-encoder for classifying 6 AD stages, from asymptomatic to clinical AD. Apart from these, support vector machines (SVM) have been widely used for AD diagnosis, as seen in [11]–[13]. So far, most of these studies focused on the clinical AD and only a handful of them examined the pre-symptom stage of AD using FNC and machine learning [8]–[10]. Moreover, the temporal dynamism of FNC has not been extensively studied, and only a limited number of works explored dFNC for AD study in conjunction with machine learning [13], [14]. Additionally, common machine learning methods may encounter challenges when confronted with intricate spatiotemporal connectomics information [15], leading to unsatisfactory performance.

In this work, we present a convolution-transformer-based neural network designed to predict asymptomatic AD subjects that are CSF biomarker-positive using dFNC. Our proposed method incorporates a spatio-temporal self-attention module, allowing for the modeling of intricate correlations between brain networks while capturing the temporal functional dynamics of the brain. The effectiveness of the proposed method is evaluated on two publicly available datasets. To provide a comprehensive analysis, we introduce AD and MCI subjects and further analyze the similarity between the dFNC patterns of subjects in different stages of AD. Additionally, our investigation extends to understanding how the functional dynamics of brain networks contribute to the model's predictions when discerning asymptomatic AD subjects from cognitively normal individuals.

## II. METHODOLOGY

### A. Dataset

*1) Emory Healthy Brain Study (EHBS):* We include data from the EHBS [16] which is a longitudinal cohort study of cognitively normal adults (50-75 years). A total of 362 subjects are selected with available fMRI scans, with 303 cognitive normal (CN) subjects and 59 asymptomatic AD (Asym) subjects. We use tTau and $A\beta_{42}$ as biomarkers to identify Asym from CN, where subjects with tTau/$A\beta_{42}$>0.24 are identified as Asym [3].

*2) Alzheimer's Disease Neuroimaging Initiative:* We also use the ADNI dataset to verify the model's performance and study the relationship between Asym and symptomatic AD and MCI subjects. We randomly select 59 AD subjects and 59 MCI subjects to mitigate the effect of class imbalance. Furthermore, we stratified the ADNI dataset according to the florbetapir standardized uptake value ratio (AV45 SUVR) and obtained 397 CN and 132 Asym subjects.

### B. Data Preprocessing

The fMRI data from the datasets are preprocessed using statistical parametric mapping (SPM12) software. The pipeline includes rigid body correction to correct the subject's head movement, slice-time correction to account for timing differences in slice acquisition, warping into the standard Montreal Neurological Institute (MNI) space using an echo planar imaging (EPI) template, resampling to isotropic voxels of 3mm³, and finally Gaussian smoothing.

To extract the dFNC data, we use the spatially-constrained ICA [17] with the Neuromark_fMRI_1.0 template [18] and obtain 53 spatially independent networks, which are grouped into sub-cortical (SC, 5), auditory (AUD, 2), sensorimotor (SM, 10), visual (VS, 9), cognitive control (CC, 16), default mode (DM, 7), and cerebellar (CB, 5) network domains, as in Fig. 1.

To estimate the FNC time courses, we apply a tapered window (width=10 TRs, i.e., 30 s) and compute the correlations across the 53 identified brain networks within the window. By sliding the window by step of 1 TR, we obtain a total of 246 windows. Therefore, for each subject, 246 53×53 symmetric correlation matrices are estimated.

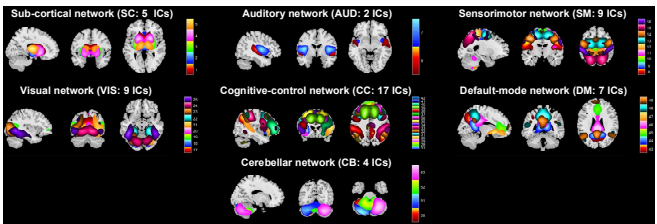

Fig. 1. Spatial maps of 53 networks assigned to 7 functional domains.

### C. Method Overview

To accurately identify Asym subjects and better understand how brain networks change during the progression AD, we take advantage of the transformer [19] and combine it with traditional convolution. Based on the backbone presented in our previous work [20], we utilize a transformer-convolution-based framework to locally and globally extract relevant features from brain network connectivity. In addition, to better capture the intricate temporal dynamics of dFNC, we propose a spatial-temporal self-attention module (as in Fig. 2) to facilitate the modeling of temporal variability between different dFNC states, thus better extracting crucial features that are significant for diagnosis.

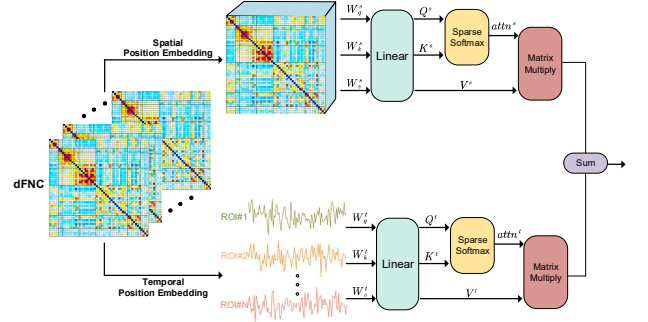

Fig. 2. The proposed spatio-temporal self-attention mechanism, which calculates spatial and temporal attention separately and fuses them to provide a spatiotemporally aware model

### D. Spatio-Temporal Self-Attention

Transformers were used for fMRI-related study in recent literature for their superb feature extracting ability [21]. The self-attention mechanism can encode long-range contextual dependencies from the data and model variations in the functional connectivity topology [22], thereby obtaining pivotal information for the designated tasks.

In this work, to better capture time-dependent aspects of spatial dependencies across dFNC states, we propose a spatio-temporal self-attention to substitute vanilla self-attention. As in Fig. 2, the proposed self-attention takes the dFNC data as input and performs position embedding in the spatial and temporal manners separately. For better performance and interpretability [20], we use a sparse function to generate the spatial self-attention $attn^s$ and temporal self-attention $attn^t$. As such, $attn^s$ represents the importance of connectives across the brain networks, while $attn^t$ denotes the importance of different dFNC states.

Standard RNN structure often overlook long-range contextual dependencies while focusing in immediate temporal links in time-series data, and while LSTM can handle this, they are typically hard to interpret [23]. The proposed temporal self-attention addresses the limitations by encoding variabilities across all dFNC states over time. By combined with spatial self-attention, it can capture higher-order representations of dynamic spatial-temporal dependencies, facilitating the asymptomatic AD diagnosis.

## III. Experimental Results

### A. Experimental Settings

We train our models based on Pytorch 2.1, with a Nvidia A100. We perform 5-fold cross-validation with 300-epoch training in each fold. We use the AdamW optimizer, with 0.05 weight decay and 0.001 learning rate. To facilitate convergence, we apply the cosine annealing rate scheduler. We use binary cross-entropy loss for binary classification and cross-entropy loss for multi-class classification.

TABLE I. COMPARE PERFORMANCES ON TWO DATASETS

| Dataset | method | acc | f1 | precision | spec | sens |
|---|---|---|---|---|---|---|
| EHBS | **proposed** | **76.76** | **45.09** | **34.21** | 75.25 | **66.10** |
|  | ResNet | 74.03 | 39.74* | 31.96* | 78.22 | 52.54* |
|  | SVM | 83.15 | 0.00* | 0.00* | 99.34 | 0.00* |
| ADNI | **proposed** | **93.76** | **86.53** | **93.81** | 80.30 | **93.76** |
|  | ResNet | 92.82* | 85.27* | 87.30* | **83.33** | 92.82* |
|  | SVM | 86.20* | 62.56* | 99.50 | 46.21* | 86.20* |

### B. Performance

We compare the proposed method with a standard residual neural network (ResNet) and a machine learning model (SVM) that is commonly applied in the literature [11]-[13]. We perform evaluation based on the EHBS dataset and the ADNI dataset. The results are shown in Table I.

As in Table I, "acc" denotes accuracy, "spec" denotes specificity (recall over the CN subjects), and "sens" denotes sensitivity (recall over the asymptomatic subjects). We further perform the paired t-test, and * in the table indicates $p<0.01$. From the table, the proposed method significantly outperforms other methods on most metrics, especially on the sensitivity, which indicates that it can more accurately diagnose subjects with asymptomatic AD. The SVM model showed inferior performance on the EHBS dataset.

### C. Analysis of AD Progression

Based on the proposed model, we explore the relationship between AD stages in terms of dFNC. Using the ADNI dataset, we perform two binary classifications: EHBS-CN vs. EHBS-Asym & ADNI-AD and EHBS-CN vs. EHBS-Asym & ADNI-MCI. As in Table II, the lower output score from the model signifies a stronger likelihood of negative class, i.e., CN.

Compared to previous results, there's a significant elevation in performance, showing a higher contrast between CN and AD/MCI than CN and Asym. In addition, Asym is similar to CN, as evidenced by the small score difference. The smaller score difference between Asym and MCI compared to Asym and AD suggests a closer similarity between MCI and Asym.

TABLE II. CONFOUND MODEL WITH AD OR MCI SUBJECTS

|  | acc | f1 | score | score on CN | score on Asym | AD/MCI |
|---|---|---|---|---|---|---|
| CN vs Asym & AD | 86.22 | 70.71 | 0.0676 | 0.1891 | 0.9814 |
| CN vs Asym & MCI | 85.99 | 72.56 | 0.0949 | 0.2599 | 0.9948 |

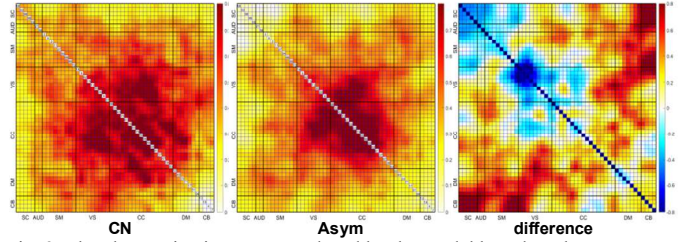

Fig. 3. The class activation maps produced by the model based on the EHBS dataset. We show the average results over CN and Asym subjects and calculate their difference (use the map of CN minus the map of Asym).

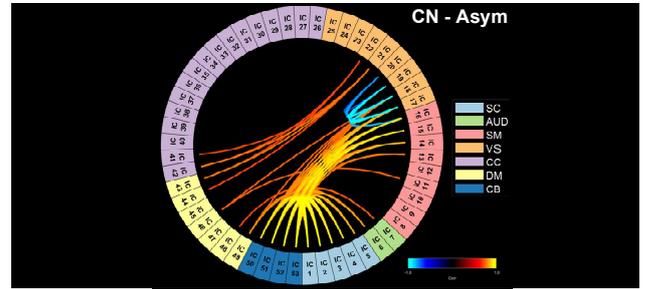

Fig. 4. Difference of correlations between brain networks of CN and Asym subjects. Note that we apply a 0.7 threshold to filter out weak correlations.

### D. Interpretable Visualizations

To show how brain networks contribute to the diagnosis, we further present the visualizations of the class activation maps produced by gradients. We evaluate several popular CAM-based methods using the confidence metric [24] and choose LayerCam [25], which has the highest average confidence. The results are shown in Fig. 3 and Fig. 4.

As in the figures, where identifying the CN or Asym subjects, the model focuses primarily on correlations within the cognitive-control network (CC) and the visual network (VS). When predicting Asym subjects, the model shows increased focus over the intra-visual domain dFNC, with decreased focus over dFNC between cerebellar-subcortical (CB-SC) and cerebellar-sensorimotor (CB-SM). From Fig. 4, the VS and SM connections have greater contributions to predicting Asym, while correlations among CB, SC, and between default-mode (DM) and VS tend to be neglected.

Our findings align with prior research. For instance, Buckley et al. [26] identified the CC network as one of the strongest predictors for Aβ change and cognitive decline. Yuan et al. [27] observed a change of connectivity in the left superior frontal gyrus and left inferior parietal lobule (components of the CC network) and the middle occipital gyrus and middle temporal gyrus (components of the VS network). Consistent results are reported in [28], presenting a significant difference between CN and Asym in CC and SM networks. Despite some previous works suggesting significant changes in the DM network [29], [30], our model indicates that it does not significantly contribute to diagnosis.

## IV. CONCLUSION

In this work, we study the pre-symptomatic stage of AD using dFNC. We introduce a novel spatial-temporal self-attention mechanism designed to capture temporal brain dynamics. We evaluate the method based on two dFNC datasets for asymptomatic AD diagnosis. We further include AD and MCI subjects to examine AD-stage relationships. Results show MCI subjects share closer similarities to asymptomatic individuals than diagnosed AD cases. Additionally, we conduct an interpretable analysis with class activations maps, highlighting specific network connections for CN and asymptomatic identification. These findings deepen the understanding of pre-symptomatic AD and offer valuable insights for future diagnostics.